\documentclass[11pt]{article} 

\usepackage[caption=false]{subfig}

\usepackage[T1]{fontenc}
\usepackage[UKenglish]{babel}
\usepackage[UKenglish]{isodate}
\usepackage[utf8]{inputenc}

\usepackage[overload]{empheq} 
\usepackage[subnum]{cases} 

\usepackage{comment}

\usepackage[margin=2cm]{geometry}

\usepackage{graphicx}
\usepackage{epstopdf}
\usepackage{pdfpages}

\usepackage{float} 

\graphicspath{{Plots/}}

\usepackage{amsmath}
\usepackage{amssymb}
\usepackage{color}
\usepackage{hyperref}
\usepackage{url}
\usepackage{amsthm}
\usepackage{booktabs}
\usepackage{bbm}
\usepackage{bm}
\usepackage{caption}
\usepackage{eucal}
\usepackage{float}
\usepackage{latexsym}
\usepackage{mathtools}
\usepackage{multicol}
\usepackage{multirow}
\usepackage{pdfpages}
\usepackage{siunitx}

\theoremstyle{plain}
\newtheorem{theorem}{Theorem}[section]

\newtheorem{lemma}[theorem]{Lemma}
\theoremstyle{example}

\newtheorem{remark}[theorem]{Remark}

\newtheorem{definition}[theorem]{Definition}
\theoremstyle{remark}

\begin{document}
	
\title{On the consistency of jump-diffusion dynamics for FX rates under inversion \footnotemark[1]}
	

\author{
	Federico Graceffa\footnotemark[2]
	\and Damiano Brigo \footnotemark[2]
	\and Andrea Pallavicini\footnotemark[3]
}

\providecommand{\email}[1]{\textit{} #1}

\renewcommand{\thefootnote}{\fnsymbol{footnote}}

\footnotetext[1]{The opinions here expressed are solely those of the authors and do not represent in any way those of their employers. \newline Corresponding author: Federico Graceffa, \texttt{federico.graceffa16@imperial.ac.uk}}

\footnotetext[2]{Department of Mathematics, Imperial College London, South Kensington Campus, London SW7 2AZ, United Kingdom.}
	
\footnotetext[3]{Department of Mathematics, Imperial College London, and Banca IMI Milan, Largo Mattioli 3, Milan MI 20121, Italy.}


\maketitle

%
%

\begin{abstract}
In this note we investigate the consistency under inversion of jump diffusion processes in the Foreign Exchange (FX) market. In other terms, if the EUR/USD FX rate follows a given type of dynamics, under which conditions will USD/EUR follow the same type of dynamics? In order to give a numerical description of this property, we first calibrate a Heston model and a SABR model to market data, plotting their smiles together with the smiles of the reciprocal processes. Secondly, we determine a suitable local volatility structure ensuring consistency. We subsequently introduce jumps and analyze both constant jump size (Poisson process) and random jump size (compound Poisson process). In the first scenario, we find that consistency is automatically satisfied, for the jump size of the inverted process is a constant as well. The second case is more delicate, since we need to make sure that the distribution of jumps in the domestic measure is the same as the distribution of jumps in the foreign measure. We determine a fairly general class of admissible densities for the jump size in the domestic measure satisfying the condition. 
\end{abstract}


\newpage

%
%

\section{Introduction}
The foreign exchange (FX) market has peculiar \emph{symmetries} that distinguish it from other markets. The first is the symmetry with respect to inversion: given an exchange rate, its reciprocal is again an exchange rate. For example, the USD-GBP is the reciprocal of the GBP-USD exchange rate. The other "symmetry" is what we could term \emph{triangular consistency} and is with respect to multiplication: given two exchange rates such that the domestic currency of one corresponds to the foreign currency of the other, their product is another exchange rate. For example, the product of USD-GBP and GBP-EUR is the cross rate USD-EUR. Triangular consistency requires that if USD-GBP and GBP-EUR are in the same model class up to reparametrization, so is USD-EUR. For an example and a related discussion with multivariate mixture models and an application involving China's FX rates see for instance Brigo et al. \cite{Brigo2019}. These two stylized facts have motivated research in understanding which mathematical models fulfil some kind of \emph{consistency conditions} which make them compatible with such empirical facts. 

Let us start with a definition
\begin{definition}
A model for $S(t)$ is said to be \emph{consistent under inversion} if the dynamics of $S(t)$ under the domestic measure is the same as the dynamics of $1/S(t)$ under the foreign measure, up to a reparametrization. Expressing $S(t)$ as a Ito stochastic differential equation (SDE), both the finite variation drift and the diffusive part will be required to have the same functional form. In case there are hidden sources of randomness, such as stochastic volatility or random jumps, consistency will be said to hold if also the description of such sources is invariant modulo reparametrization.
\end{definition}
Such a requirement can be justified from different points of view. Firstly, in principle there is no reason why an exchange rate and its inverse should be described in substantially different ways. They are actually the same entity, just seen from two different perspectives. Furthermore, in terms of design of libraries, it is helpful to have a consistent dynamics for all FX rates involved in transactions.

The issue of consistency with respect to inversion was raised, for instance, by Brigo et al. \cite{Brigo2015} in the context of multi currency CDSs and FX rate devaluation in conjunction with default events. In Section 2.2. the authors explain that when pricing quanto CDS one might be interested in pricing either under the liquid-currency measure or the contractual-currency measure. FX symmetry plays a role in that the measure change affects all risk factors whose dynamics is defined under a measure different from the one in which they were calibrated. 

The Heston model is certainly one of the most widespread \cite{heston1993closed, rouah2013heston}. Its consistency with respect to inversion was first addressed by Del Ba{\~n}o Rollin \cite{Rollin2008}, who showed that the Heston model is indeed well behaving (see also \cite{de2013smiles,gnoatto2017coherent}). On the other hand, inconsistent models are numerous: for instance, the Garch stochastic volatility model \cite{gnoatto2017coherent}, the SABR model \cite{gnoatto2017coherent,de2013smiles}, the Hull-White stochastic volatility model \cite{gnoatto2017coherent}, and the Scott model \cite{de2013smiles}. By following the \emph{intrinsic currency} framework introduced by Doust \cite{Doust2007,doust2012stochastic}, De Col et al. \cite{de2013smiles} presented a multi-factor SV model of Heston type which remains invariant under a risk-neutral measure change. This approach was later generalized by Gnoatto \cite{gnoatto2017coherent}, who introduced a consistent affine stochastic volatility model. The intrinsic currency approach was employed by Gnoatto and Grasselli as well \cite{gnoatto2014affine}, who extended the model presented previously in \cite{de2013smiles} to the case where the stochastic factors driving the volatilities of the exchange rates belong to the cone of positive semidefinite $d \times d$ matrices $S_d^{+}$. Specifically, they showed that their model is at the same time an affine multifactor stochastic volatility model for the FX rate where the instantaneous variance is driven by a Wishart process, and a Wishart affine short-rate model. Recently, Graceffa et al. \cite{Graceffa2019} studied consistency with respect to inversion of fairly general classes of local stochastic volatility (LSV) models, determining general conditions that a LSV model has to satisfy in order to be consistent. Finally, it is worth mentioning that this problem was also discussed in the context of semimartingales, see for instance the works of Eberlein and Papapantoleon \cite{eberlein2005symmetries} and and Eberlein et al. \cite{eberlein2008duality} who discussed the so-called duality principle.

This paper aims at including jumps into the analysis and discussing how Poisson and compound Poisson processes behave under inversion in the FX rate. It is structured as follows. In Section \ref{GeneralModel_numerics} we present a fairly general jump-diffusion model with local-stochastic volatility, and illustrate some numerical results highlighting the consistency and inconsistency of the Heston model and SABR model respectively. In section \ref{InvLocalStructure} we discuss consistency for a general local volatility structure, identifying some suitable functional forms satisfying the required property. In section \ref{InvConstantJumpSize} we focus without loss of generality on the jump component, and discuss the case where jump size is constant. Here consistency with respect to inversion turns out to be automatically satisfied, since the jump size is a constant as well. Finally, in Section \ref{InvCompoundPoisson} we analyze the more complicated case of compound Poisson processes. We identify a fairly general class of jump size distributions which are invariant, up to a reparametrization, under the transformation from domestic to foreign measure.

%
%

\section{General model and numerics}\label{GeneralModel_numerics}
Consider the jump-diffusion with local stochastic model
\begin{subequations}\label{stochVolJumps}
	\begin{align}
	d S(t) & = \Delta r S(t) dt + \eta(t,V(t)) \sigma(t,S(t)) S(t) d W_1^{\mathbb{Q}_d}(t) +  S(t_{-}) J^d d N^{\mathbb{Q}_d}(t) \\
	d V(t) & = m(t,V(t)) dt + \xi(t,V(t)) d W_2^{\mathbb{Q}_d}(t)
	\end{align}
\end{subequations}
where the process $S(t)$ denotes the exchange rate, $\Delta r = r^d - r^f$ the differential of the domestic and foreign risk free interest rates, $m,\eta,\sigma,\xi: [0,T] \to \mathbb{R}$ measurable functions, $\mathbb{Q}_d$ denotes the risk neutral domestic measure, $W_1^{\mathbb{Q}_d}(t),W_2^{\mathbb{Q}_d}(t)$ standard Brownian motions under the domestic measure, $N^{\mathbb{Q}_d}(t)$ a Poisson process under the domestic measure with intensity $\lambda^d$, and $J^d$ the size of the relative jump of the exchange rate. The Poisson process will be assumed to be independent from the Brownian motions while the two Brownian motions will in general be correlated.

As mentioned in the introduction, a model $S(t)$ is said to be consistent under inversion if the SDE describing $S$ in the domestic measure and the SDE describing $1/S$ in the foreign measure are the same, up to a reparametrization. Furthermore, any hidden source of randomness, such as stochastic volatility or stochastic jumps, must be described by the same kind of SDE/distribution. To give a numerical measure of the inconsistency, we can consider, for the sake of simplicity, the Heston model (which is consistent) and the SABR model (which is not) \cite{Graceffa2019}. The Heston model is
\begin{subequations}
	\begin{align}
	d S(t) & = \Delta r S(t) dt + \sqrt{V(t)} S(t) d W_{1}^{\mathbb{Q}_d}(t) \\
	d V(t) & = \kappa (\bar{V} - V(t)) dt + \sigma \sqrt{V(t)} d W_{2}^{\mathbb{Q}_d}(t).
	\end{align}
\end{subequations}
This model is well known to be consistent. Its inverse is 
\begin{subequations}
	\begin{align}
	d Y(t) & = - \Delta r Y(t) dt + \sqrt{V(t)} Y(t) d W_{1}^{\mathbb{Q}_f}(t) \\
	d V(t) & = (\kappa - \rho \sigma) \left(\frac{\kappa}{\kappa - \rho \sigma}\bar{V} - V(t) \right) dt + \sigma \sqrt{V(t)} d W_{2}^{\mathbb{Q}_f}(t).
	\end{align}
\end{subequations}
Indeed, the model dynamics followed by the inverse FX rate is again of Heston type. It is important to point out, though, that the term $k - \rho \sigma$ should be positive. The reason being that, otherwise, the volatility model is not mean reverting anymore. Such a condition is easily fulfilled in case, for example, the correlation between the asset and volatility processes is negative.

The SABR model reads as
\begin{subequations}
	\begin{align}
	d S(t) & = \Delta r S(t) dt + S^\beta(t) d W_{1}^{\mathbb{Q}_d}(t) \\
	d V(t) & = \nu V(t) d W_{2}^{\mathbb{Q}_d}(t).
	\end{align}
\end{subequations}
Unlike Heston, the SABR model is inconsistent. The inverse of the SABR is
\begin{subequations}
	\begin{align}
	d Y(t) & = - \Delta r Y(t) dt + v(t) Y^{2-\beta}(t) d W_{1}^{\mathbb{Q}_f}(t) \\
	d V(t) & = \nu \rho Y^{1-\beta}(t) V^2(t) dt + \nu V(t) d W_{2}^{\mathbb{Q}_f}(t).
	\end{align}
\end{subequations}
The model dynamics is indeed not a SABR model anymore. In order to provide the reader with a clearer understanding of what consistency means in practice, we fit both models and their inverses to market data. As we shall see, consistency will imply that the smile of a model and the smile of the inverse model will match almost perfectly. Using an inconsistent model, instead, will cause the two smiles to be markedly different.

In the first plot we calibrate Heston model and SABR model to the market data, and show the resulting smiles. 
\begin{figure}[H]
	\centering
	\includegraphics[scale=.75]{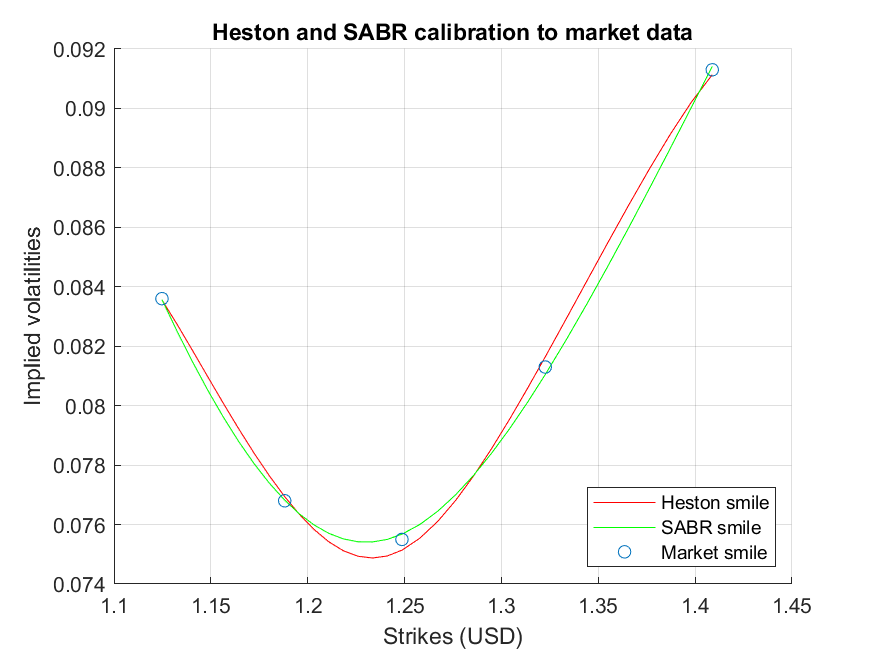}
	\caption{Calibration Heston and SABR model to EUR/USD market volatility, $30$-th January 2018, $3$ months maturity.}
	\label{}
\end{figure}
The underlying asset is the EUR/USD exchange rate as of $30$-th January 2018, with spot $S_0 = 1.24122$. We used five market volatilities: $10$-delta put, $25$-delta put, ATM, $25$-delta call, $10$-delta call. The maturity is $T = 3$ months. The scattered strikes are those from market data, and were computed via (see \cite{Clark2011}, Eq. (3.8))
\begin{equation}
K_{\text{market}} := F_{0,T} \exp \left(\frac{1}{2} \sigma^2_{ATM} T\right),
\end{equation}
$F_{0,T}$ denoting the current forward price. In our specific case, $ F_{0,T} = 1.2478$ and $\sigma_{ATM} = 0.0755$. As usual, calibration was carried out my minimizing the sum of squared differences between market volatilities and model implied volatilities. In both cases, results are quite satisfactory. Heston calibration gives the parameters
\begin{equation}
v^*_0 = 0.0025, \quad, \theta^* = 0.0287, \quad k^* = 1.1718, \quad \sigma^* = 0.1720, \quad \rho^* = 0.0952.
\label{HestonPars}
\end{equation}
The condition ensuring consistency is indeed satisfied. On the other hand, calibrating the SABR model gives us the parameters
\begin{equation}
\alpha^* = 0.0748, \quad \rho^* = 0.1435, \quad \nu^* = 0.7330,
\end{equation}
with $\alpha_{shift} = 9.8986 \times 10^{-8}$, where, we recall, $\alpha$ denotes the current SABR volatility and it is shifted so as to match the ATM volatility, and $\nu$ denotes the volatility of volatility. The parameter $\beta$, instead, is chosen a priori to be $0.50$. In order to make Heston and SABR smiles look smooth, we built a denser strikes vector and then performed a spline interpolation. Next, we illustrate the consistency of the Heston model. Using the calibrated parameters \eqref{HestonPars} of the Heston model, we priced options on the reciprocal exchange rate using the reciprocal of the Heston model, which is known to be a Heston model as well. Then, given these prices, we employed a standard numerical routine and obtained the corresponding implied volatilities, coherent with the Heston reciprocal. Finally, we plotted the smile of these implied volatilities against the one of the original model.

Visualizing both smiles in the same plot, we see that they are almost overlapping. More precisely, the norm of the difference between the two is of order $10^{-14}$. Such an overlap indicates the model and its reciprocal have the same volatility, that is, the volatility in the models is described by the same kind of stochastic dynamics. This, in turn, is a  clear sign confirming model consistency.
\begin{figure}[H]
	\centering
	\includegraphics[scale=.75]{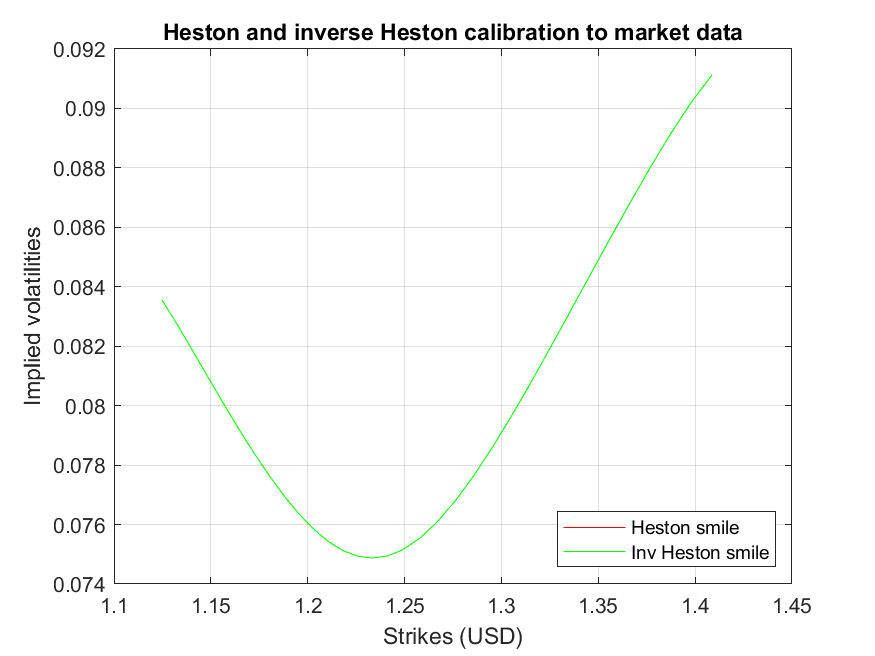}
	\caption{Consistency Heston model with respect to inversion: Heston smile and inverse smile are almost exactly overlapping.}
	\label{}
\end{figure}
Repeating the same procedure with the SABR model and its reciprocal, we see that the result is considerably different.
\begin{figure}[H]
	\centering
	\includegraphics[scale=.75]{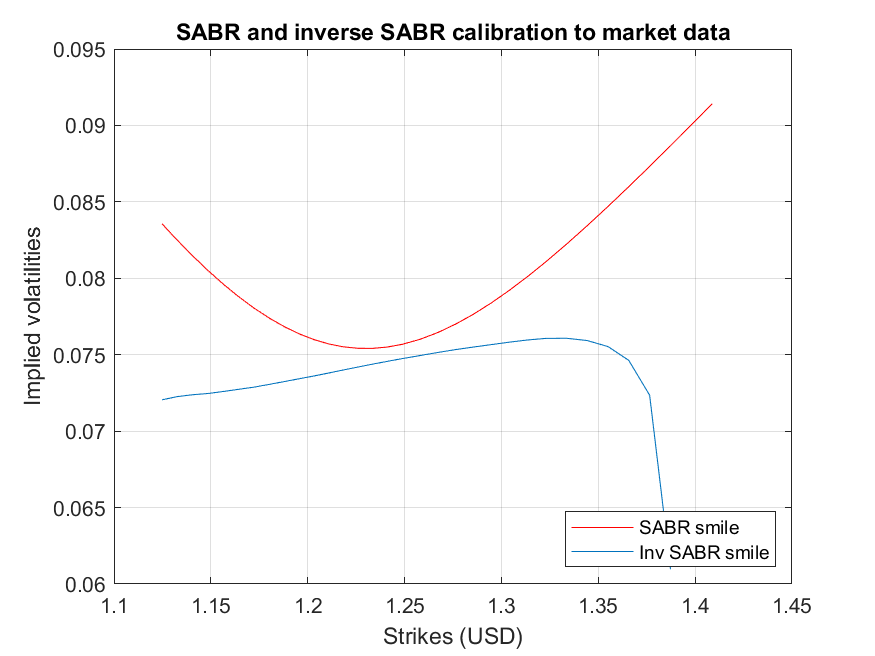}
	\caption{Inconsistency SABR model with respect to inversion}
	\label{}
\end{figure}
Unlike what happens in Figure 2, the two smiles do not overlap at all. This clearly suggests this model is not consistent under inversion.

%
%

\section{Inversion with local volatility structure}\label{InvLocalStructure}
After showing a few numerical examples, we proceed by analysing more complex models. We start by investigating the scenario where no jumps nor stochastic volatility are present, that is we specify the model \eqref{stochVolJumps} with $J^d = 0$ and $\eta = 1$. Inversion of local stochastic volatility was discussed in \cite{Graceffa2019}. The authors determined an affine condition for the local volatility component, and a relationship linking the functions $m,xi,\eta$. Here we propose a further viable specification for the general volatility. Let us consider the model
\begin{equation}
d S(t) = \Delta r S(t) dt + \sigma(S,t) S(t) d W^{\mathbb{Q}_d}(t).
\end{equation}
Then, by Ito's formula, the inverted dynamics in the domestic measure reads as
\begin{equation}
d \left( \frac{1}{S(t)} \right) = \left[-\frac{1}{S(t)} \Delta r + \sigma^2(S,t) \frac{1}{S(t)} \right] dt - \frac{1}{S(t)} \sigma(S,t) d W^{\mathbb{Q}_d}(t).
\end{equation}
Implementing the change of measure from domestic to foreign 
\begin{equation}
d W^{\mathbb{Q}_f}(t) = d W^{\mathbb{Q}_d}(t) - \sigma(S(t),t) dt,
\end{equation}
implies
\begin{equation}
d \left( \frac{1}{S(t)} \right) = - \Delta r \frac{1}{S(t)} dt - \frac{1}{S(t)} \sigma(S(t),t) d W^{\mathbb{Q}_f}(t).
\end{equation}
Therefore, the dynamics of the inverted exchange rate $Y(t):=1/S(t)$ in the foreign measure becomes
\begin{equation}
d Y(t) = - \Delta r Y(t) dt - \sigma \left(\frac{1}{Y(t)},t \right) Y(t) d W^{\mathbb{Q}_f}(t).
\end{equation}
This means that in order to ensure consistency, we will require
\begin{equation}
\sigma \left(\frac{1}{Y},t \right) \sim \sigma(Y,t),
\label{FunctionalFormLocVol}
\end{equation}
where by $\sim$ we mean the same functional form. A non trivial function satisfying this is, for example, the logarithm, since
\begin{equation}
\log \left( \frac{1}{x} \right) = - \log(x).
\end{equation}
More generally, we might consider any \emph{polynomial of logarithms}. Another class is given by
\begin{equation}
\sigma(x,t) = \frac{1}{x^k} + \frac{1}{x^{k-1}} + \dots + 1 + x^{k-1} + x^k.  
\end{equation}
Interestingly, this last expression is a local volatility which is useful in practice thanks to its flexibility in the parametrization. Generally, we could consider
\begin{equation}
\sigma(x) = f(\log(x)),
\end{equation}
with function $f$ having some sort of symmetry around the $x$-axis
\begin{equation}
f(x) \sim f(-x).
\end{equation}

%
%

\section{Inversion of jump diffusion with constant jump size}\label{InvConstantJumpSize}
By virtue of the independence of the Brownian motion from the Poisson process, we can, without loss of generality, set the volatility structure to be constant, and focus on the jump component of our model. In the current section we will assume the jump size to be constant,
\begin{subequations}
	\begin{align}
	S(t) & = S(t_{-}) + \Delta S(t) = S(t_{-}) + S(t_{-}) \gamma^d \\
	& = S(t_{-}) (1+\gamma^d).
	\end{align}
\end{subequations}
Specifying $\eta = 1, \sigma(S(t),t) = \sigma$ in \eqref{stochVolJumps}, our model becomes
\begin{equation}
d S(t) = (\Delta r - \gamma^d \lambda^d) S(t) dt + \sigma S(t) d W^{\mathbb{Q}_d}(t) + S(t_{-}) \gamma^d d N^{\mathbb{Q}_d}(t),
\label{DynamicsS}
\end{equation}
In order to determine whether, and under which conditions, consistency is fulfilled, we define, as above, the inverse exchange rate $Y(t) := \frac{1}{S(t)}$. Applying Ito's formula for jump-diffusion processes (see Cont and Tankov \cite{Cont2004}, Prop 8.14,) yields
\begin{equation}
d Y(t) = (\Delta r + \gamma^d \lambda^d + \sigma^2) Y(t) dt - \sigma Y(t) d W^{\mathbb{Q}_d}(t) + Y(t_{-}) \left(-\frac{\gamma^d}{1 + \gamma^d} \right) d N^{\mathbb{Q}_d}(t).
\end{equation}
Next, we perform a change of measure so as to express $Y$ in the foreign measure. As it is well known (see e.g. Brigo and Mercurio \cite{brigo2007interest}), the change of measure is defined via the Radon-Nikodym derivative
\begin{equation}
L(t) := \frac{S(t) B^f(t)}{S(0) B^d(t)}, 
\label{RNDerivative}
\end{equation}
$B^d(t), B^f(t)$ denoting the domestic and foreign bank accounts respectively.  In general, this can be rewritten in closed form as (see e.g. Shreve \cite{shreve2004stochastic})
\begin{equation}
L(t) = L_1(t) L_2(t)
\label{RNDerivative_factor}
\end{equation}
with
\begin{equation}
L_1(t) = \exp \left\{\sigma W(t) - \frac{1}{2} \sigma^2 t \right\}
\end{equation}
responsible for the Brownian motion and 
\begin{equation}
L_2(t) = e^{(\lambda^f - \lambda^d) t} \left( \frac{\lambda^f}{\lambda^d} \right)^{N(t)}
\label{ChangePoisson}
\end{equation}
responsible for the Poisson process, $\lambda^f$ denoting the intensity of the Poisson process in the foreign measure. In differential form, we might write
\begin{equation}
d L_1(t) = \sigma L_1(t) d W^{\mathbb{Q}^d}(t) 
\end{equation}
\begin{equation}
d L_2(t) = \frac{\lambda^f - \lambda^d}{\lambda^d} L_2(t) d M^{\mathbb{Q}^d}(t) = \gamma^d L_2(t)  d M^{\mathbb{Q}^d}(t),
\end{equation}
with $d M^{\mathbb{Q}^d}(t):=d N^{\mathbb{Q}^d}(t) - \lambda dt $ a martingale. In the second expression, the first equality is due to \eqref{ChangePoisson}, while the second is due to \eqref{RNDerivative}. More compactly, applying formula \eqref{RNDerivative_factor} and noting that Brownian motion and Poisson process are independent:
\begin{equation}
d L(t) = \sigma L(t) d W^{\mathbb{Q}^d}(t) + \gamma^d L(t)  d M^{\mathbb{Q}^d}(t).
\end{equation}
Hence, we deduce that
\begin{equation}
\frac{\lambda^f - \lambda^d}{\lambda^d} = \gamma^d,
\end{equation}
that is
\begin{equation}
\lambda^f = \lambda^d (1+\lambda^d).
\label{NewLambda}
\end{equation}
The new Brownian motion is given by
\begin{equation}
d W^{\mathbb{Q}_f}(t) := d W^{\mathbb{Q}_d}(t) - \sigma dt.
\end{equation}

\begin{remark}
It could be interesting to notice that equation \eqref{NewLambda} can be deduced heuristically as follows (see also \cite{bjork2011introduction})
\begin{subequations}
	\begin{align}
	\lambda^f dt & = \lambda^d dt + \frac{\mathbb{E}[d L(t) d N(t) | \mathcal{F}_t]}{L(t)} \\ 
	& = \lambda^d dt + \frac{\mathbb{E}[ \gamma^d L(t) (d N(t))^2 | \mathcal{F}_t ] }{L(t)} \\
	& = \lambda^d dt + \gamma^d \lambda dt \\
	& = \lambda^d (1+\gamma^d) dt.
	\end{align}
\end{subequations}
\end{remark}
Let us now notice that there appear to be two choices for the definition of the new jump size $\gamma^f$: we can define it as the whole term multiplying the Poisson process or that term with a minus in front. We opt for the first choice, that is 
\begin{equation}
\gamma^f := - \frac{\gamma^d}{1 + \gamma^d}.
\end{equation}
The reason for doing so is that in this way both the two jumps sizes in the different measures have domain $D = (-1,+\infty)$. Indeed, 
\begin{subequations}
	\begin{align}
	\gamma^d \to -1^{+} & \implies \gamma^f \to + \infty \\
	\gamma^d \to + \infty & \implies \gamma^f \to -1^{+}.
	\end{align}
\end{subequations}
Therefore, the dynamics of $Y$ under the foreign measure reads as
\begin{equation}
d Y(t) = Y(t) (\Delta r - \gamma^f \lambda^f) dt - \sigma Y(t) d W^{\mathbb{Q}^f}(t) + Y(t_{-}) \gamma^f d N^{\mathbb{Q}^f}(t).
\end{equation}
Since $\gamma^d$ is constant, so is $\gamma^f$. Hence, consistency is readily fulfilled. It is also easy to check that $Y$ is correctly compensated. This happens when
\begin{equation}
- \lambda^d \gamma^d +  \lambda^f \gamma^f = 0
\end{equation}
and this is satisfed in view of \eqref{NewLambda} and the definition of $\gamma^f$. In absolute values,
\begin{equation}
\frac{\lambda^f}{\lambda^d} = \left| \frac{\gamma^d}{\gamma^f} \right|.
\end{equation}
This means that the higher the jump size in the domestic measure, the higher the jump frequency in the foreign measure. Since the foreign jump size is decreasing as a function of the domestic jump size, the foreign intensity must somehow compensate this effect and then increase. In other words, \emph{the Poisson process in the foreign measure is expected to have a higher number jumps, but with a lower size}. 

%
%

\section{Inversion of jump diffusion with compound Poisson process}\label{InvCompoundPoisson}
Finally, we discuss the case where the jump size is random, that is when the exchange rate is driven by a compound Poisson process. Consistency will now be more restrictive, as we will require the distribution of jump sizes not to be affected by the measure change. The aim of this section will be to determine a fairly general class of densities for the jump size in the domestic measure for which such condition will be satisfied.
\begin{equation}
d S(t) = (\Delta r - \beta^d \lambda^d) S(t) dt + \sigma S(t) d W^{\mathbb{Q}_d}(t) + S(t_{-}) d K^{\mathbb{Q}_d}(t),
\label{DynamicsSCompound}
\end{equation}
where $K^{\mathbb{Q}_d}(t)$ is a compound Poisson process (under the domestic measure)
\begin{equation}
K^{\mathbb{Q}_d}(t) := \sum_{i=1}^{N^{\mathbb{Q}_d}(t)} J_i^d,
\end{equation}
with the jump sizes $J_i^d$ are i.i.d. (independent of the processes $W$ and $N$) and
\begin{equation}
\beta^d := \mathbb{E}^d[J^d]
\end{equation}
is the expectation of the domestic jump size under the domestic measure. The jump part might be conveniently rewritten as
\begin{subequations}
	\begin{align}
	d K^{\mathbb{Q}_d}(t) & =  K^{\mathbb{Q}_d}(t + dt) - K^{\mathbb{Q}_d}(t) \\
	& = \sum_{i=1}^{N^{\mathbb{Q}_d}_{t+dt}} J_i^d - \sum_{i=1}^{N^{\mathbb{Q}_d}_t} J_i^d \\
	& = \sum_{i=1}^{N^{\mathbb{Q}_d}_t + d N^{\mathbb{Q}_d}_t} J_i^d - \sum_{j=1}^{N^{\mathbb{Q}_d}_t} J_i^d \\
	& = J_{1}^d d N^{\mathbb{Q}_d}_t.
	\end{align}
\end{subequations}
Hence, our model can we rewritten also as
\begin{equation}
d S(t) = (\Delta r - \beta^d \lambda^d) S(t) dt + \sigma S(t) d W^{\mathbb{Q}_d}(t) + S(t_{-}) J_{1}^d d N^{\mathbb{Q}_d}(t).
\label{DynamicsSCompound2}
\end{equation}
For the sake of clarity, we remark that the compensator $\beta^d \lambda^d$ guarantees absence of arbitrage. Since $S(t-)$ is $\mathcal{F}_{t-}$-measurable, 
\begin{equation}
\mathbb{E}^d_{t-} \left[ S(t-) d K^{\mathbb{Q}_d}(t) \right] = S(t-) \mathbb{E}^d_{t-} \left[d K^{\mathbb{Q}_d}(t) \right] = S(t-) \mathbb{E}^d[J^d] \lambda^d dt.
\end{equation}
In close form, this is (see Shreve \cite{shreve2004stochastic})
\begin{equation}
S(t) = S(0) \exp \left\{ \sigma W^{\mathbb{Q}_d}(t) + (\Delta r - \beta^d \lambda^d - \frac{1}{2} \sigma^2 )t \right\} \prod_{i=1}^{N^{\mathbb{Q}_d}(t)} (J_i^d + 1).
\end{equation}
We can readily see that the domain $D$ of the density of the jump size must be contained in $(-1,+\infty)$. Performing the inversion and changing measure yields
\begin{equation}
d Y(t) = (-\Delta r + \beta^d \lambda^d) Y(t) dt - \sigma Y(t) d W^{\mathbb{Q}_f}(t) + Y(t) \left(-\frac{J^d}{1+J^d}\right) d N^{\mathbb{Q}_f}(t).
\end{equation}
Analogously to the constant scenario, we might define
\begin{equation}
J^f := - \frac{J^d}{1 +J^d}.
\end{equation}
In order to determine expressions for $\lambda^f$ and $f^f$, we look at the Radon-Nykodim derivative
\begin{equation}
L(t) := \frac{S(t) B^f(t)}{S(0) B^d(t)}.
\end{equation}
Its differential is
\begin{equation}
d L(t) = \sigma L(t) d W^{\mathbb{Q}_d}(t) + L(t) d M^{\mathbb{Q}_d}(t),
\end{equation}
with $M$ being a martingale defined via
\begin{equation}
d M(t)^{\mathbb{Q}_d} := d K^{\mathbb{Q}_d}(t) - \beta^d \lambda^d dt.
\end{equation}
In general, the RD derivative describing the joint change of measure of a Brownian  motion and compound Poisson process is (see Shreve \cite{shreve2004stochastic})
\begin{equation}
L(t) = L_1(t) L_2(t),
\end{equation}
with
\begin{equation}
L_1(t) = \exp \left\{ \sigma W(t) - \frac{\sigma^2}{2} t \right\}
\end{equation}
and
\begin{equation}
L_2(t) = e^{(\lambda^f - \lambda^d)t} \prod_{i=1}^{N(t)} \frac{\lambda^f}{\lambda^d} \frac{f^f(J_i)}{f^d(J_i)}.
\end{equation}
In differential form, we have (see Shreve \cite{shreve2004stochastic})
\begin{equation}
d L_2(t) = L_2(t-) d (H(t) - \lambda^f t) - L_2(t-) d (N(t)-\lambda^d t)
\end{equation}
with 
\begin{equation}
H(t) = \sum_{i=1}^{N(t)} \frac{\lambda^f}{\lambda} \frac{f^f(J_i)}{f^d(J_i)}.
\end{equation}
Also,
\begin{equation}
d H = \Delta H(t) = \frac{\lambda^f}{\lambda^d} \frac{f^f(Y_1)}{f^d(Y_1)} \Delta N(t).
\end{equation}
Therefore,
\begin{subequations}
	\begin{align}
	d L_2(t) & = L_2(t) \left[ d H(t) - \lambda^f dt - d N(t) + \lambda^d dt \right] \\
	& = L_2(t) \left[ \left(\frac{\lambda^f}{\lambda^d} \frac{f^f(Y_1)}{f^d(Y_1)}  - 1 \right) d N(t) + (\lambda^d - \lambda^f) dt\right].
	\end{align}
\end{subequations}
In our context of FX measure change, we have
\begin{equation}
L(t) = \exp \left\{ \sigma W^{\mathbb{Q}_d}(t) + (-\beta^d \lambda^d - \frac{1}{2} \sigma^2) t \right\} \prod_{i=1}^{N^{\mathbb{Q}_d}(t)} (Y_i + 1).
\end{equation}
The differential of the jump part can be written as
\begin{equation}
d L_2(t) = L_2(t) \left[ J_1^d d N^{\mathbb{Q}_d} - \beta^d \lambda^d dt\right].
\end{equation}
It is now evident that, comparing the $dt$ and $dN$ terms, we obtain
\begin{equation}
\lambda^d - \lambda^f = - \beta^d \lambda^d,
\end{equation}
that is, the intensity in the foreign measure is
\begin{equation}
\lambda^f = (1+\beta^d) \lambda^d.
\end{equation}
Moreover, 
\begin{equation}
\frac{\lambda^f}{\lambda^d} \frac{f^f(J_1)}{f^d(J_1)} - 1 = J_1,
\end{equation}
that is
\begin{equation}
\frac{f^f(J_1)}{f^d(J_1)} = (1+J_1) \frac{\lambda^d}{\lambda^f}.
\end{equation}
Therefore, the probability distribution function of the jump size in the foreign measure is
\begin{equation}
f^f(x) = f^d(x) (1+x) \frac{\lambda^d}{\lambda^f}.
\label{Relation_ff_fd}
\end{equation}
For the sake of completeness, we check that the inverted process $Y(t)$ is free of arbitrage. This is true if and only if
\begin{equation}
\mathbb{E}^d[J^d] \lambda^d + \mathbb{E}^f \left[-\frac{J^d}{1+J^d}\right] \lambda^f = 0.
\end{equation}
In terms of densities, this is equivalent to
\begin{equation}
\lambda^d \int_{-1}^{+\infty} x f^d(x) dx + \lambda^f \int_{-1}^{+\infty} - \frac{x}{1+x} f^f(x) dx = 0.
\label{NoarbCompoundPoisson}
\end{equation}
and this is readily satisfied in view of \eqref{Relation_ff_fd}.

As far as consistency is concerned, we require \emph{the density $f^d$ of the jump size $J^d$ under the domestic measure to belong to the same class as the density of $J^f$ under the foreign measure}. Notice that $f^f$ is the density of $J^d$ under the foreign measure.
\begin{remark}
For small $J$, at first order we have
\begin{equation}
-\frac{J}{1+J} = \frac{1}{1+J} -1 \approx 1 - J - 1 = - J,
\end{equation}
that is
\begin{equation}
J^f = - J^d.
\end{equation}
This means that for small jumps, consistency is automatically satisfied. 
\end{remark}

\begin{lemma}
Let $X$ be a generic random variable and define
\begin{equation}
Y = - \frac{X}{1+X}.
\end{equation}
Let $f_X$ be the density of $X$ and $f_Y$ the density of $Y$. Then, for any measure $\mathbb{P}$,
\begin{equation}
f_Y(y) = f_X \left(-\frac{y}{1+y} \right) \frac{1}{(1+y)^2}.
\end{equation}
\begin{proof}
We have
\begin{subequations}
	\begin{align}
	\mathbb{P}(Y \leq y) & =  \mathbb{P} \left(- \frac{X}{1+X} \leq y \right) \\
	& = \mathbb{P} (-X \leq y + X y) \\
	& = \mathbb{P} \left(X(1+y) \geq -y \right) \\
	& = \mathbb{P} \left(X \geq - \frac{y}{y+1} \right) \\
	& = 1 - \mathbb{P} \left(X < - \frac{y}{y+1} \right).
	\end{align}
\end{subequations}
So,
\begin{equation}
F_Y(y) = 1 - F_X \left(- \frac{y}{1+y} \right).
\end{equation}
Differentiating, we conclude.
\end{proof}
\end{lemma}
Let us denote by $f^f_{J^d}$ the density of $J^d$ under the foreign measure $\mathbb{Q}_f$, by $f^d_{J^d}$ the density of $J^d$ under the domestic measure $\mathbb{Q}_d$, and by $f^f_{J^f}$ the density of $J^f$ under the foreign measure $\mathbb{Q}_f$. Then, in view of equation \eqref{Relation_ff_fd}, we write
\begin{equation}
f^f_{J^d}(x) = f^d_{J^d}(x) (1+x) \frac{\lambda^d}{\lambda^f}.
\end{equation}
Moreover, in light of the lemma above, it holds
\begin{equation}
f^f_{J^f}(y) = f^f_{J^d} \left(-\frac{y}{1+y} \right) \frac{1}{(1+y)^2}.
\end{equation}
Combining the two equations yields the relationship
\begin{equation}
f^f_{J^f}(y) = f^d_{J^d} \left(-\frac{y}{1+y} \right) \frac{1}{(1+y)^3} \frac{\lambda^d}{\lambda^f}.
\end{equation}
At this point we might ask ourselves which kind of densities are appropriate. Keep in mind the domain $D = (-1,+\infty)$. Let us consider a power law distribution with cutoff function
\begin{equation}
f^d_{J^d}(x) = c \frac{1}{(1+x)^\alpha} e^{g(x)},
\end{equation}
where $c$ is the normalization constant. Then,
\begin{subequations}
	\begin{align}
	f^f_{J^f}(y) & = c  \frac{1}{\left(1-\frac{y}{1+y}\right)^{\alpha}} e^{g \left(-\frac{y}{1+y} \right)} \frac{1}{(1+y)^3} \frac{\lambda^d}{\lambda^f} \\
	& = c \frac{\lambda^d}{\lambda^f} e^{g \left(-\frac{y}{1+y}\right)} \frac{1}{(1+y)^{3-\alpha}}.
	\end{align}
\end{subequations}
We might define the new scaling parameter $\beta:=3-\alpha$. Consistency is then fulfilled only for those cut-off functions such that 
\begin{equation}
g(y) \sim g \left( -\frac{y}{1+y}\right).
\end{equation}
Notice that in this way the cut-off function ensures convergence at both $-1$ and $+\inf$, since
\begin{equation}
\lim_{x \to +\infty} g(x) = \lim_{x \to -1^+} g(x) = -\infty.
\end{equation}
A possible guess for $g$ is
\begin{equation}
g(x) = - q \frac{x^2}{1+x},
\end{equation}
with $q$ positive constant. Indeed,
\begin{equation}
\frac{-\left(\frac{y}{1} \right)^2}{1-\frac{y}{1+y}} = - \frac{y^2}{(1+y)^2} (1+y) = - \frac{y^2}{1+y}.
\end{equation}
Summing up, a good candidate for the jump size density is
\begin{equation}
f^d_{J^d}(x) = c \frac{1}{(1+x)^\alpha} e^{- q \frac{x^2}{1+x}},
\end{equation}
This density is defined on the interval $[-1,+\infty)$, and it is made of two components: a power law part depending on the scaling parameter $\alpha$, and an exponential cutoff depending on a positive parameter $q$.

%
%

\section{Conclusion}
In this paper we discussed the consistency of jump-diffusion processes with respect to their reciprocal processes, and therefore their applicability in the FX market. After presenting the general model and illustrating numerically the Heston consistency and SABR inconsistency, we presented general classes of local volatility models fulfilling our conditions. Then, we discussed inversion of a jump process with constant jump size. In this simple case, consistency is automatically verified, since the jump size of the inverted process, defined under the foreign measure, is a constant as well. More involved is the inversion of a compound Poisson process. In this case, we determined a fairly general admissible density for the jump size in the domestic measure ensuring that the jumps distribution of the original process under the domestic measure is, up to a reparametrization, the same as the jumps distribution of the inverted process under the foreign one.

\newpage

\bibliography{references}
\bibliographystyle{siam}

\end{document}